\documentclass{elsart3p}

\usepackage{amssymb}
\usepackage{amsmath}
\usepackage{graphicx}

\def\ecart{\noalign{\medskip}}

\begin{document}

\begin{frontmatter}

\title{Unified model for small-$t$ and high-$t$ scattering at high energies: predictions at RHIC and LHC}

\author[russe]{E. Martynov},
\author[fr]{B.~Nicolescu\corauthref{cor}}
\corauth[cor]{Coresponding author.}
\ead{nicolesc@lpnhe.in2p3.fr}

\address[russe]{N.N. Bogolyubov Institute for Theoretical Physics, National Academy of Sciences of Ukraine, Kiev, Ukraine}
\address[fr]{Theory Group,  Laboratoire de Physique Nucl\'eaire  et des Hautes  \'Energies (LPNHE)\thanksref{lpnhe},   }
\address{CNRS and Universit\'e Pierre et Marie Curie, Paris}

\thanks[lpnhe]{ Unit\'e de Recherche des Universit\'es
Paris 6 et Paris 7, associ\'ee au CNRS}


\begin{abstract}
The urgency of predictions in large-$t$ region at LHC stimulated us to present a unified model of small and high $t$ scattering at high energies. Our model is based upon a safe theoretical ground: analyticity, unitarity, Regge behavior, gluon exchange and saturation of bounds established in axiomatic quantum field theory. We make precise predictions for the behavior of the differential cross sections at high $t$, the evolution of the dip-shoulder structure localized in the region $0.5\lesssim |t| \lesssim 0.8 \mbox{ GeV}^{2}$ and the radical violation of the exponential behavior of the first diffraction cone at small $t$.
\end{abstract}

\begin{keyword}
Large-t scattering; high energies; LHC; RHIC; Odderon; Pomeron
\PACS 12.40.Nn; 13.75.Cs; 13.85.Dz; 13.85.Lg

\end{keyword}

\end{frontmatter}


R.F.~Avila, P.~Gauron and B.~Nicolescu (AGN) recently published a good fit of forward and non-forward data in $pp$ and $\bar pp$ scattering, leading to intriguing predictions at RHIC and LHC \cite{AGN}.

The AGN model was defined to be valid only at moderate-$t$ values ($0\le\vert t\vert\le 2.6$ GeV$^2$). We stress that this model  is based upon a safe theoretical ground: analyticity, unitarity, Regge behavior and saturation of bounds established in axiomatic quantum field theory.

In the present paper we generalize this model by extending its validity till high-$t$ values
($0\le\vert t\vert\le 16$ GeV$^2$) and we present our predictions at RHIC and LHC.
Such predictions are urgently needed in view of the near start of experiments at LHC.

As data we used the dataset recently proposed by J.R.~Cudell, A.~Lengyel and E.~Martynov (CLM), who built a coherent dataset of all existing data for
$5\le \sqrt{s} \le 1800$ GeV and $0\le\vert t\vert\le 16$ GeV$^2$\cite{CLM}.
The raw data are, of course, contained in the Durham Database \cite{HEPdata}, but CLM introduced a detailed study of the systematic errors and gathered into a common format more than 260 subsets of data from more than 80 experimental papers.
The CLM can be considered as a \textit{standard dataset}, which is available online \cite{StandDataset}.

Let us first define the even-under-crossing and the odd-under-crossing amplitudes $F_\pm(s,t)$:
\begin{equation}
\label{eq:1}
F_{\pm}=\frac{1}{2}(F_{pp}(s,t)\pm F_{\bar pp}(s,t))\ ,
\end{equation}
which are normalized so that
\begin{equation}
\begin{array}{l}
\label{eq:2}
\sigma_{t}(s)=\frac{1}{\sqrt{s(s-4m_{p}^{2})}}Im F(s,t=0),\\ 
\rho(s)=\frac{Re F(s,t=0)}{Im F(s,t=0)},
\end{array}
\end{equation}
\begin{equation}
\begin{array}{l}
\label{eq:3}
\frac{d\sigma}{dt}(s,t)=\frac{1}{16\pi gs(s-4m_{p}^{2})}|F(s,t)|^{2}\ , \\ 
g=0.3893797 \mbox{ mb\!}\cdot \mbox{\!GeV}^{-2}\ ,
\end{array}
\end{equation}
$m_p$ being the mass of the proton.

In the AGN model\footnote{Notation for the signature factors differs from the one of the original AGN model.} $F_+(s,t)$ is defined as a superposition of the following contributions: the Froissaron $F_+^H(s,t)$, the Pomeron pole $F_+^P(s,t)$, the Pomeron-Pomeron Regge cut $F_+^{PP}(s,t)$, the secondary $f_2-a_2$ reggeon $F_+^R(s,t)$ and the reggeon-Pomeron cut $F_+^{RP}(s,t)$:
\begin{equation}
\begin{array}{rl}
\label{eq:4}
F_+(s,t) =  & F_+^H(s,t)+F_+^P(s,t)+F_+^{PP}(s,t)+\\ 
                  & F_+^R(s,t)+F_+^{RP}(s,t)\ ,
\end{array}
\end{equation}
where
\begin{equation}
\label{eq:5}
\begin{array}{l}
F_+^H(s,t)=is\left\{\right.
H_{1}\ln^{2}\tilde s \displaystyle \frac{2J_{1}(K_{+} \tilde \tau)}{K_{+}\tilde \tau}\exp(b_{+1}t)\\
\ecart
+H_{2}\ln\tilde s J_{0}(K_{+} \tilde \tau)\ln\tilde s\exp(b_{+2}t)\\
\ecart
+H_{3}[J_{0}(K_{+} \tilde \tau)-K_{+} \tilde \tau J_{1}(K_+ \tilde \tau)]\exp(b_{+3}t)\bigr\}
\end{array}
\end{equation}
\begin{multline}
\label{eq:6}
F_{+}^{P}(s,t)=C_{P}\exp(b_{P}t)\Phi_{+}(\alpha_{P}(t))\\
(s/s_{0})^{\alpha_{P}(t)}\ ,
\end{multline}
\begin{multline}
\label{eq:7}
F_{+}^{PP}(s,t)=\frac{C_{PP}}{\ln\tilde s}\exp(b_{PP}t)\Phi_{+}(\alpha_{PP}(t))\\
(s/s_{0})^{\alpha_{PP}(t)},
\end{multline}
\begin{multline}
\label{eq:8}
F_{+}^{R}(s,t)=C_{R}^{+}\exp(b_{R}^{+}t)\Phi_{+}(\alpha_{R}^{+}(t))\\
(s/s_{0})^{\alpha_{R}^{+}(t)},
\end{multline}
\begin{multline}
\label{eq:9}
F_{+}^{RP}(s,t)=\frac{-itC_{RP}^{+}}{\ln \tilde s}\exp(b_{RP}^{+}t) \Phi_{+}(\alpha_{RP}^{+}(t))\\
(s/s_{0})^{\alpha_{RP}^{+}(t)},
\end{multline}
\begin{equation}
\label{eq:10}
\tilde s=(s/s_{0})\exp \left(-\frac{1}{2}i\pi\right ), \quad s_{0}=1{\rm GeV}^{2},
\end{equation}
\begin{equation}
\label{eq:11}
\tilde \tau =\sqrt{-t/t_{0}}\ln\tilde s, \quad t_{0}=1{\rm GeV}^{2},
\end{equation}
\begin{equation}
\label{eq:12}
\Phi_{+}(\alpha(t))=i\sin\left (\frac{\pi}{2}\alpha(t)\right )-
\cos\left (\frac{\pi}{2}\alpha(t)\right ),
\end{equation}
\begin{align}
\ &\alpha_{P}(t)=1+\alpha_{P}'t\ , \\
& \alpha_{PP}(t)=1+\alpha_{P}'t/2\ ,\\
&\alpha_{R}^{+}(t)=\alpha_{R}^{+}(0)+\alpha_{R}^{+'}t\ , \\
& \alpha_{RP}^{+}(t)=
\alpha_{R}^{+}(0)+\frac{\alpha_{P}'\alpha_{R}^{+'}}{\alpha_{P}'+\alpha_{R}^{+'}}t\ ,
\end{align}

$H_1,\ H_2,\ H_3,\ K_+,\ b_{+1},\ b_{+2},\ b_{+3}, C_{P},\ b_{P},\ C_{PP},$ $b_{PP},\ C_{R}^{+},\ b_{RP}^{+},\ C_{RP}^{+},\ b_{RP}^{+},\ \alpha_{P}',\ \alpha_{R}^{+}(0),\ \alpha_{R}^{+'}$  being constants.
$J_0$ and $J_1$ are Bessel functions.

In its turn, the $F_-(s,t)$ is defined, in the AGN model, as a superposition of the Maximal Odderon Contribution $F_-^{Mo}(s,t)$, the Odderon pole $F_-^0(s,t)$, the Odderon-Pomeron cut $F_-^{OP}(s,t)$, the secondary $\rho-\omega$ reggeon $F_-^R(s,t)$ and the reggeon-Pomeron cut $F_-^{RP}(s,t)$:
\begin{equation}
\begin{array}{rl}
\label{eq:15}
F_{-}(s,t)= & F_{-}^{MO}+F_{-}^{O}(s,t)+F_{-}^{OP}(s,t)+\\
& F_{-}^{R}(s,t)+F_{-}^{RP}(s,t)\ ,
\end{array}
\end{equation}
where
\begin{equation}
\label{eq:16}
\begin{array}{l}
F_{-}^{MO}(s,t)=s\left \{\right.
O_{1}\ln^{2}\tilde s \displaystyle \frac{\sin(K_{-} \tilde \tau)}{K_{-}\tilde \tau}\exp(b_{-1}t)\\
\ecart
+O_{2}\ln\tilde s  \cos(K_{-} \tilde \tau)\exp(b_{-2}t)\\
+O_{3}\exp(b_{-3}t)\left.\right \}.
\end{array}
\end{equation}
\begin{multline}
\label{eq:17}
F_{-}^{O}(s,t)=C_{O}\exp(b_{O}t)\Phi_{-}(\alpha_{O}(t))\\
(s/s_{0})^{\alpha_{O}(t)}\ ,
\end{multline}
\begin{multline}
\label{eq:18}
F_{-}^{OP}(s,t)=\frac{C_{OP}}{\ln \tilde s}\exp(b_{OP}t)\Phi_{-}(\alpha _{OP}(t))\\
(s/s_{0})^{\alpha_{OP}(t)}\ ,
\end{multline}
\begin{multline}
\label{eq:19}
F_{-}^{R}(s,t)=-C_{R}\exp(b_{R}^{-}t)\Phi_{-}(\alpha_{R}^{-}(t))\\
(s/s_{0})^{\alpha_{R}^{-}(t)}\ ,
\end{multline}
\begin{multline}
\label{eq:20}
F_{-}^{RP}(s,t)=\frac{-tC_{RP}}{\ln \tilde s}\exp(b_{RP}^{-}t) \Phi_{-}(\alpha_{RP}^{-}(t))\\
(s/s_{0})^{\alpha_{RP}^{-}(t)}\ ,
\end{multline}
\begin{equation}
\label{eq:21}
\Phi_{-}(\alpha(t))=i\cos\left (\frac{\pi}{2}\alpha(t)\right )+
\sin\left (\frac{\pi}{2}\alpha(t)\right )
\end{equation}
\begin{align}
\ &\alpha_{O}(t)=1+\alpha_{O}'t\ ,\\
&\alpha_{OP}(t)=1+\frac{\alpha_{P}'\alpha_{O}'}{\alpha_{P}'+\alpha_{O}'}\ t\ , \\
&\alpha_{R}^{-}(t)=\alpha_{R}^{-}(0)+\alpha_{R}^{-'}t\ , \\
&\alpha_{RP}^-(t)=
\alpha_R^-(0)+\frac{\alpha'_P\alpha_R^{-'}}{\alpha'_P+\alpha_R^{-'}}\ t\ \label{eq:new},
\end{align}
$O_1,\ O_2,\ O_3,\ K_-,\ b_1^-,\  b_2^-,\  b_3^-,\ C_O,\ b_O,\ C_{OP}$, $b_{OP},\ C_R^-,\
b_R^-,\ C_{RP}^-,\ b_{RP}^-,\ \alpha_0^{\prime},\ \alpha_R^-(0),\ \alpha_R^{- \prime }$
being constants.

The AGN amplitudes defined through eqs.~(\ref{eq:1})-({\ref{eq:new}), lead to an excellent description of the standard dataset:
\begin{equation}
\label{eq:23}
\chi^2/dof=1.16
\end{equation}
for a total of 1966 experimental points.
The $\sigma_T$ and $\rho$ data (238 experimental points) correspond to an energy range 5 GeV $\le\sqrt{s}\le$ 1.8 TeV.
The $d\sigma/dt$ data (1728 experimental points) correspond to an energy range 9.3 GeV $\le \sqrt{s}\le$ 1.8 TeV and to a $t$-range 0.1 $\le \vert t\vert\le$ 2.6 GeV$^2$.
The presence of the Maximal Odderon is, of course, crucial in order to describe the experimental difference between the $pp$ and $\bar pp$ differential cross-sections at
$\sqrt{s}=5.8$ GeV in the dip-shoulder region 1.1  $< \vert t \vert < $
1.5 GeV$^2$ \cite{ISR53}.

{\footnotesize
\begin{table}[t]
\caption{\newline The values of parameters in $F_+(s,t)$}
\centering
\begin{tabular}{|cccc|}
\hline
     Name    & Dimension       &  Value & Error    \\
\hline
$H_{1}$            & ${\rm mb\cdot GeV}^{2}$ &  0.22  &  0.002     \\
$H_{2}$            & ${\rm mb\cdot GeV}^{2}$ & -0.016  &  0.001       \\
$H_{3}$            & ${\rm mb\cdot GeV}^{2}$ &  27.43  &  0.30                  \\
$K_{+}$            &                         &  0.20  &  0.001      \\
$C_{p}$            & ${\rm mb\cdot GeV}^{2}$ &  2.65  &  0.16    \\
$C_{PP}$           & ${\rm mb\cdot GeV}^{2}$ & -40.04  &  2.18   \\
$C_{R}^{+}$        & ${\rm mb\cdot GeV}^{2}$ &  77.61  &  1.71    \\
$C_{RP}^{+}$       & ${\rm mb\cdot GeV}^{2}$ & -23.73  &  0.70    \\
$\alpha_{R}^{+}(0)$&                         &  0.67  &  0.01    \\
$\alpha_{R}^{+'}$   & ${\rm GeV}^{-2}$       &  0.80  &  0.01    \\
$\alpha_{P}' $     & ${\rm GeV}^{-2}$        &  0.41  &  0.01                  \\
$b_{+1} $          & ${\rm GeV}^{-2}$        &  3.06  &  0.03    \\
$b_{+2} $          & ${\rm GeV}^{-2}$        &  0.66  &  0.01    \\
$b_{+3} $          & ${\rm GeV}^{-2}$        &  4.57  &  0.03    \\
$b_{P} $           & ${\rm GeV}^{-2}$        &  0.31  &  0.07    \\
$b_{PP} $          & ${\rm GeV}^{-2}$        &  9.18  &  0.44    \\
$b_{R}^{+} $       & ${\rm GeV}^{-2}$        &  4.80  &  0.14   \\
$b_{RP}^{+} $      & ${\rm GeV}^{-2}$        &  0.46  &  0.02    \\
$N_{+}$            & ${\rm mb\cdot GeV}^{2}$ & -9.42  &  0.33    \\
$t_{+} $           & ${\rm GeV}^{2}$         &  0.45 &  0.01     \\
\hline
\end{tabular}
\end{table}
}
{\footnotesize
\begin{table}[t]
\caption{\newline The values of parameters in $F_-(s,t)$}
\centering
\begin{tabular}{|cccc|}
\hline
     Name    & Dimension       &  Value & Error    \\
\hline
 $O_{1}$            & ${\rm mb\cdot GeV}^{2}$ &  0.002  &  00.001  \\
  $O_{2}$            & ${\rm mb\cdot GeV}^{2}$ &  0.12  &  0.01  \\
 $O_{3}$            & ${\rm mb\cdot GeV}^{2}$ &  -2.13 &  0.08  \\
 $K_{-}$            &                         &  0.011  & 0.002  \\
$C_{O}$            & ${\rm mb\cdot GeV}^{2}$ &  0.88  &  0.02  \\
$C_{OP}$           & ${\rm mb\cdot GeV}^{2}$ &  0.061  & 0.005  \\
$C_{R}^{-}$        & ${\rm mb\cdot GeV}^{2}$ &  38.00  &  0.37  \\
$C_{RP}^{-}$       & ${\rm mb\cdot GeV}^{2}$ &  -4753 &  101  \\
$\alpha_{R}^{-}(0)$&                         &  0.53 &  0.01  \\
 $\alpha_{R}^{-'} $ & ${\rm GeV}^{-2}$        &  0.80  &  0.01  \\
 $\alpha_{O}' $     & ${\rm GeV}^{-2}$        &  0.10  &  0.01 \\
$b_{-1} $          & ${\rm GeV}^{-2}$        &  1.07  &  0.02  \\
$b_{-2} $          & ${\rm GeV}^{-2}$        &  2.87  &  0.06  \\
$b_{-3} $          & ${\rm GeV}^{-2}$        &  3.09  &  0.04  \\
$b_{O} $           & ${\rm GeV}^{-2}$        &  2.29  &  0.02  \\
$b_{OP} $          & ${\rm GeV}^{-2}$        &  0.18  &  0.01  \\
$b_{R}^{-} $       & ${\rm GeV}^{-2}$        &  1.42  &  0.05  \\
$b_{RP}^{-} $      & ${\rm GeV}^{-2}$        &  8.44  &  0.77  \\
$N_{-}$            & ${\rm mb\cdot GeV}^{2}$ &  2.16  &  0.20  \\
$t_{-} $           & ${\rm GeV}^{2}$         &  0.61  &  0.01  \\
 $A_{MO} $          & ${\rm GeV}^{-2}$        &  7.32  &  0.32  \\
$A_{O} $           & ${\rm GeV}^{-2}$        &  8.87  &  0.41  \\
 $t_{c} $           & ${\rm GeV}^{2}$         & -0.16  &  0.004  \\
\hline
\end{tabular}
\end{table}
}

Now, we generalize the AGN model by introducing the following three ingredients:\\
1.
  Contributions $N_\pm(s,t)$ to $F_\pm(s,t)$ behaving like $t^{-4}$ at high $t$:
 \begin{equation}
\label{eq:24}
N_{+}(s,t)=isN_{+}\ln\tilde s \frac{-t}{(1-t/t_{+})^{5}}\ ,
\end{equation}
\begin{equation}
\label{eq:25}
N_{-}(s,t)=sN_{-}\ln\tilde s \frac{-t}{(1-t/t_{-})^{5}}\ .
\end{equation}
These additional terms add just 4 free parameters more as compared with the original AGN model. Similar forms as those of Eqs.~(\ref{eq:24})-(\ref{eq:25}) were used in Ref.~\cite{DL}.

The experimental $t^{-4}$ behavior at large $t$ is well known.

The theoretical motivation for $N_-(s,t)$ was given longtime ago by Donnachie and Landshoff: the Odderon 3-gluon exchange, with elementary gluons \cite{Donnachie:1990wd}.
The motivation for $N_+(s,t)$ is more obscure, but it could be correlated with the $C=+$ part of the exchange of 3 elementary gluons.\\
2.
  A crossover factor $Z_-(t)$ in the $F_-(s,t)$ amplitude
  \begin{equation}
\label{eq:26}
Z_-(t)=\frac{\tanh(1-t/t_{c})}{\tanh(1)}.
\end{equation}
This factor introduce just one more free parameter as compared with the original AGN model.

The phenomenological motivation for such a $Z_-(t)$ factor is obvious: the $\bar pp$ and $pp$ cross at very small $t$ ($\vert t\vert\simeq 0.16$ GeV$^2$) and the respective crossover point is practically independent on $s$.\\
3.
  Two linear functions $(1+A_{MO}t)$ and $(1+A_Ot)$ which multiply the Maximal Odderon and the Odderon pole contributions, respectively.
  These factors have also a phenomenological justification: they allow us to better describe the smallness of the Odderon forward couplings at present energies.
  These linear functions introduce just 2 more free parameters as compared with the original AGN model.

  In fact we considered a big number of fits by taking into account just one additional ingredient or a combination of two additional ingredients.
  The net result is that, for an excellent fit of the standard dataset we need all the 3 ingredients.
  This result is not trivial, because we add a ridiculously small number of free parameters (7 free parameters) as compared with the total of 36 free parameters of the AGN model.
  Moreover, we considerably increase the number of fitted data points: from 1966 to 2892.
  Therefore, we are convinced that the 3 ingredients represent a safe theoretical generalization of the AGN model at high $t$.

\begin{figure}[t]
\begin{center}
\includegraphics[scale=0.4]{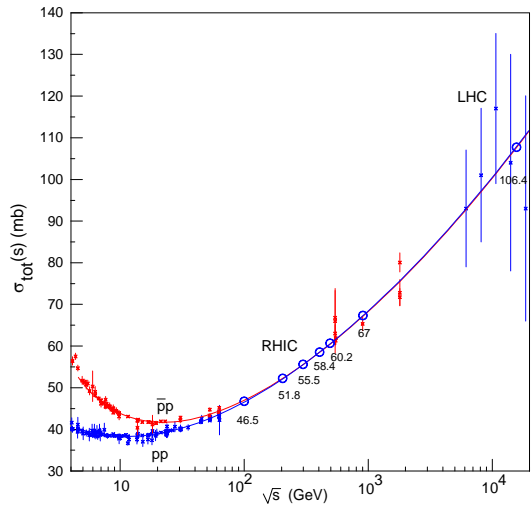}
\caption{Total cross-sections. Our predictions at RHIC and LHC are also shown.}
\label{fig1}
\end{center}
\end{figure}

\begin{figure}[t]
\begin{center}
\includegraphics[scale=0.4]{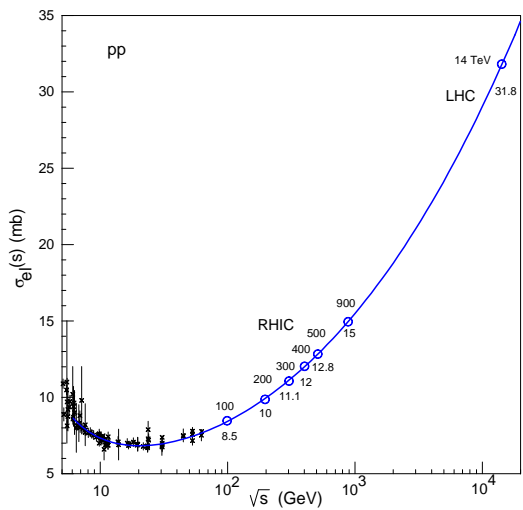}
\caption{$pp$ elastic cross-sections. Our predictions at RHIC and LHC are also shown.}
\label{fig2}
\end{center}
\end{figure}

The best value of the free parameters involved in the amplitudes defined in Eqs.~(\ref{eq:4})-(\ref{eq:new}) and (\ref{eq:24})-(\ref{eq:26}) with the additional 3 ingredients are obtained through a $\chi^2$ MINUIT minimization and they are shown in Tables 1 and 2.


The $\chi^2/dof$ is surprisingly small:
\begin{equation}
\label{eq:27}
\chi^2/dof=1.23,
\end{equation}
by taking into account the huge number of experimental data which are fitted (2892 points) in a huge range of energies
\begin{equation}
\label{eq:28}
5\le \sqrt{s}\le 1800 \mbox{ TeV}
\end{equation}
and in a huge range of $t$\footnote{The range in Eq.~(\ref{eq:29}) is justified in Ref.~\cite{CLM}. We exclude 244 points, most of them at low energies, from the dataset in Ref.~\cite{CLM} (corresponding to 5 $pp$ sets and 2 $\bar pp$ sets), because they are in conflict with all the other data.}
\begin{equation}
\label{eq:29}
t=0 \quad \mbox{ and }\quad  0.1\le\vert t\vert\le 16 \mbox{ GeV}^2
\end{equation}
The fit presented in this paper is certainly the best existing model for describing the data.
We present here all our analytic forms and all the values of the parameters so any author can compare his/her own model with our model.

The quality of the fit is shown in Fig.~1 (total cross-sections $\sigma_{tot}(s)$), Fig.~2 (elastic cross-sections $\sigma_{el}(s)$), Fig.~3 ($\rho$-parameter $\rho(s)$), Fig.~4 ($pp$ differential cross-sections), and Fig.~5 ($\bar pp$ differential cross-sections).
In Figs. 1-3 we also show our predictions for $\sigma_{tot}(s),\ \sigma_{el}(s)$ and $\rho(s)$ at RHIC and LHC energies.

Our predictions for the $pp$ differential cross-sections at RHIC and LHC energies are shown in Fig.~6.
We predict (see Fig.~6) a dip - shoulder structure persisting at RHIC and LHC energies, namely: a dip at $\vert t\vert \simeq 0.7$ GeV$^2$ for $\sqrt{s}=500$ GeV, which moves to $\vert t\vert\simeq 0.5$ GeV$^2$ for $\sqrt{s}=14$ TeV, while the shoulder present at $\vert t\vert\simeq 1$ GeV$^2$ for $\sqrt{s}=500$ GeV moves to $\vert t\vert\simeq 0.8$ GeV$^2$ at $\sqrt{s}=14$ TeV.

In Fig.~7, we plot the slope $B_{pp}(s,t)$ of the differential cross-sections at RHIC and LHC energies, as a function of $s$ \textit{and} $t$.
A very interesting effect can be contemplated in Fig.~7: at RHIC energies $B(s,t)$ is constant at fixed $s$ in a quite large range of $t$ but at LHC energies the constancy of $B(s,t)$ is totally lost: the exponential behavior of $d\sigma/dt$ is no more valid.

All these predictions can be verified at RHIC and LHC energies. In particular, the TOTEM experiment will explore the large $|t|$ $pp$ elastic scattering till 8 GeV$^{2}$ \cite{Eggert2006}. Also experiments at RHIC will be very helpful. The verification of our predictions can help to clarify the dynamics involved in small and high $t$ scattering.


\newpage

\begin{figure}[t]
\begin{center}
\includegraphics[scale=0.4]{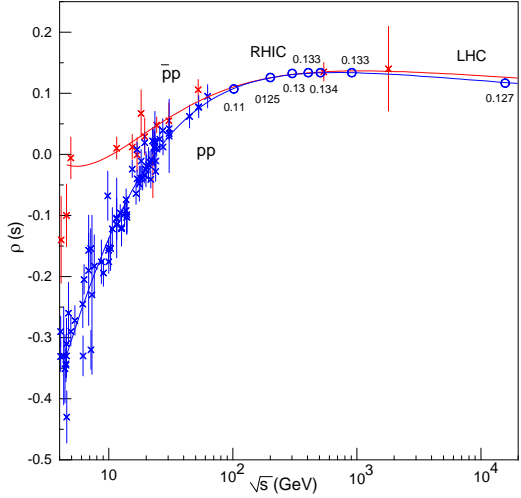}
\caption{$\rho$-parameter. Our predictions for $pp$ scattering at RHIC and LHC are also shown.}
\label{fig3}
\end{center}
\end{figure}

\begin{figure}[t]
\begin{center}
\includegraphics[scale=0.4]{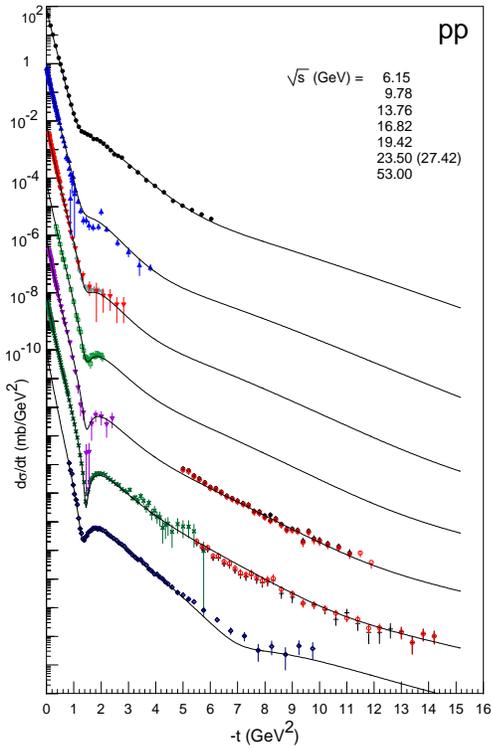}
\caption{$pp$ differential cross-sections in the energy range $6.15\le\sqrt{s}\le 53$ GeV.
Data and theoretical values are multiplied by $10^{-2(n-1)}$, where $n$ is the number of curve (and corresponding dataset) starting from the top.}
\label{fig4}
\end{center}
\end{figure}

\begin{figure}[b]
\begin{center}
\includegraphics[scale=0.4]{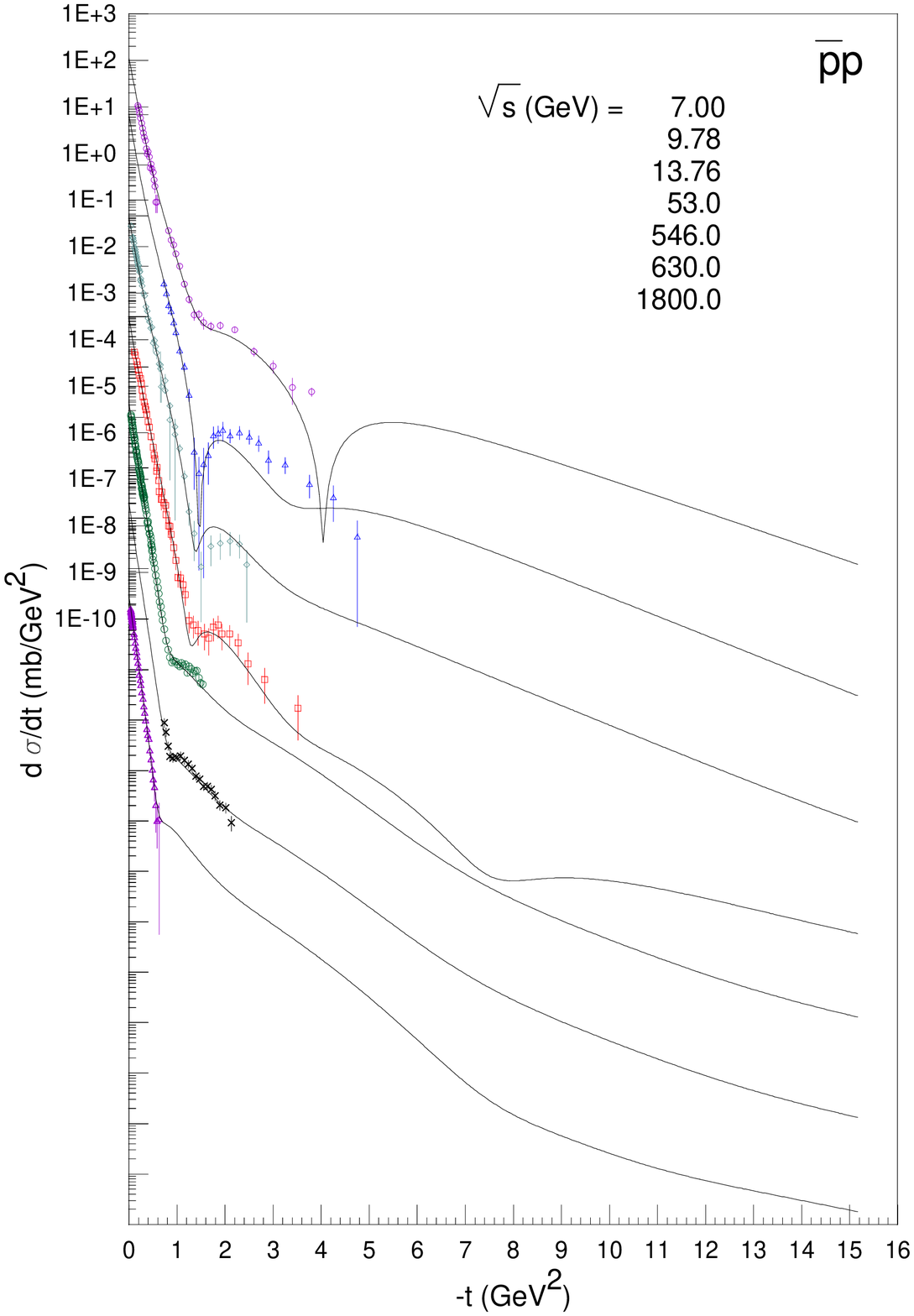}
\caption{$\bar pp$ differential cross-sections in the energy range $7\mbox{ GeV }\le\sqrt{s}\le 1.8$ TeV.
Data and theoretical values are multiplied by $10^{-2(n-1)}$, where $n$ is the number of curve (and corresponding dataset) starting from the top.}
\label{fig5}
\end{center}
\end{figure}

\begin{figure}[b]
\begin{center}
\includegraphics[scale=0.4]{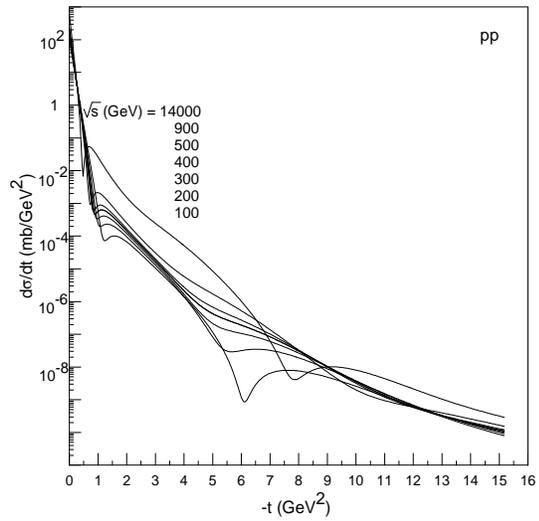}
\caption{Evolution of $pp$ differential cross-sections from RHIC to LHC energies.}
\label{fig6}
\end{center}
\end{figure}

\clearpage

\begin{figure}
\begin{center}
\includegraphics[scale=0.4]{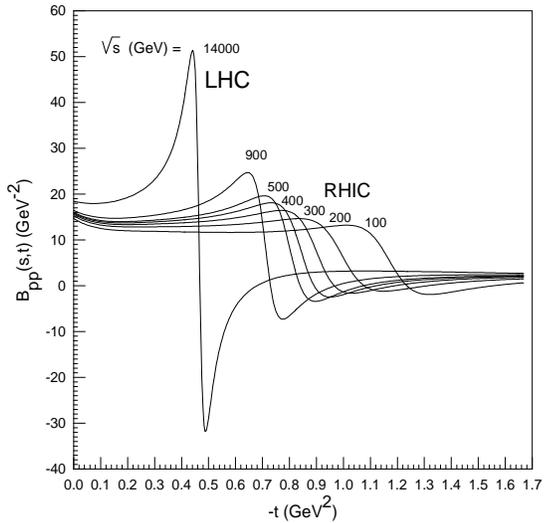}
\caption{Evolution of the slope in $pp$ scattering
$B_{pp}(s,t)\equiv \dfrac{1}{t}\mathrm{ln}\left(\dfrac{d\sigma/dt}{d\sigma/dt\vert_{t=0}}\right)$  
from RHIC to LHC energies.}
\label{fig7}
\end{center}
\end{figure}

\newpage

\section*{Acknowledgments} 
We thank Prof. J.~R.~Cudell and Dr.~Pierre Gauron for useful discussions.
One of us (E.M.) thanks Prof. Pascal Debu for the kind hospitality at LPNHE Paris, where the present work was finished. 
E.M. also thanks the Ukrainian Fund of Fundamental Researches and the Department of Physics and Astronomy of National Academy of Sciences of Ukraine for a support of this work partially performed in Kiev.

\end{document}